\documentclass[dvips]{acta}
\usepackage{supertabular,lscape,epsfig}
\usepackage{amssymb}
\usepackage{amsmath}
\usepackage{float}
\mathchardef\mhyphen="2D
\usepackage{placeins}
\DeclareSymbolFont{ppa}{OT1}{ppl}{m}{it}
\DeclareMathSymbol{\vv}{\mathalpha}{ppa}{'166}

\SetPages{315}{339}

\SetVol{68}{2018}

\begin{document}
\newcommand\pvalue{\mathop{p\mhyphen {\rm value}}}
\newcommand{\TabApp}[2]{\begin{center}\parbox[t]{#1}{\centerline{
  {\bf Appendix}}
  \vskip2mm
  \centerline{\small {\spaceskip 2pt plus 1pt minus 1pt T a b l e}
  \refstepcounter{table}\thetable}
  \vskip2mm
  \centerline{\footnotesize #2}}
  \vskip3mm
\end{center}}

\newcommand{\TabCapp}[2]{\begin{center}\parbox[t]{#1}{\centerline{
  \small {\spaceskip 2pt plus 1pt minus 1pt T a b l e}
  \refstepcounter{table}\thetable}
  \vskip2mm
  \centerline{\footnotesize #2}}
  \vskip3mm
\end{center}}

\newcommand{\TTabCap}[3]{\begin{center}\parbox[t]{#1}{\centerline{
  \small {\spaceskip 2pt plus 1pt minus 1pt T a b l e}
  \refstepcounter{table}\thetable}
  \vskip2mm
  \centerline{\footnotesize #2}
  \centerline{\footnotesize #3}}
  \vskip1mm
\end{center}}

\newcommand{\MakeTableApp}[4]{\begin{table}[p]\TabApp{#2}{#3}
  \begin{center} \TableFont \begin{tabular}{#1} #4 
  \end{tabular}\end{center}\end{table}}

\newcommand{\MakeTableSepp}[4]{\begin{table}[p]\TabCapp{#2}{#3}
  \begin{center} \TableFont \begin{tabular}{#1} #4 
  \end{tabular}\end{center}\end{table}}

\newcommand{\MakeTableee}[4]{\begin{table}[htb]\TabCapp{#2}{#3}
  \begin{center} \TableFont \begin{tabular}{#1} #4
  \end{tabular}\end{center}\end{table}}

\newcommand{\MakeTablee}[5]{\begin{table}[htb]\TTabCap{#2}{#3}{#4}
  \begin{center} \TableFont \begin{tabular}{#1} #5 
  \end{tabular}\end{center}\end{table}}

\newcommand{\MakeTableH}[4]{\begin{table}[H]\TabCap{#2}{#3}
  \begin{center} \TableFont \begin{tabular}{#1} #4 
  \end{tabular}\end{center}\end{table}}

\newcommand{\MakeTableHH}[4]{\begin{table}[H]\TabCapp{#2}{#3}
  \begin{center} \TableFont \begin{tabular}{#1} #4 
  \end{tabular}\end{center}\end{table}}
\newfont{\bb}{ptmbi8t at 12pt}
\newfont{\bbb}{cmbxti10}
\newfont{\bbbb}{cmbxti10 at 9pt}
\newcommand{\uprule}{\rule{0pt}{2.5ex}}
\newcommand{\douprule}{\rule[-2ex]{0pt}{4.5ex}}
\newcommand{\dorule}{\rule[-2ex]{0pt}{2ex}}
\def\thefootnote{\fnsymbol{footnote}}
\begin{Titlepage}

\Title{OGLE Collection of Galactic Cepheids\footnote{Based on observations
obtained with the 1.3-m Warsaw telescope at the Las Campanas Observatory of
the Carnegie Institution for Science.}}
\Author{
A.~~U~d~a~l~s~k~i$^1$,~~
I.~~S~o~s~z~y~ñ~s~k~i$^1$,~~
P.~~P~i~e~t~r~u~k~o~w~i~c~z$^1$,~~
M.\,K.~~S~z~y~m~a~ñ~s~k~i$^1$,\\
D.\,M.~~S~k~o~w~r~o~n$\!^1$,~~
J.~~S~k~o~w~r~o~n$\!^1$,~~
P.~~M~r~ó~z$\!^1$,~~
R.~~P~o~l~e~s~k~i$\!^2$,~~
S.~~K~o~z~³~o~w~s~k~i$\!^1$,\\
K.~~U~l~a~c~z~y~k$^3$,~~
K.~~R~y~b~i~c~k~i$^1$,~~
P.~~I~w~a~n~e~k$^1$~~
and~~M.~~W~r~o~n~a$^1$
}
{$^1$Warsaw University Observatory, Al.~Ujazdowskie~4, 00-478~Warszawa, Poland\\
e-mail: udalski@astrouw.edu.pl\\
$^2$Department of Astronomy, Ohio State University, 140 W. 18th Ave., Columbus, OH~43210, USA\\
$^3$Department of Physics, University of Warwick, Gibbet Hill Road, Coventry, CV4~7AL,~UK}
\Received{November 4, 2018}
\end{Titlepage}

\Abstract{We present here a new major part of the OGLE Collection of
Variable Stars -- OGLE Collection of Galactic Cepheids. The new dataset
was extracted from the Galaxy Variability Survey images -- a dedicated
large-scale survey of the Galactic disk and outer bulge conducted by the
OGLE project since 2013.

The OGLE collection contains 2721 Cepheids of all types -- classical,
type~II and anomalous. It more than doubles the number of known Galactic
classical Cepheids. Due to the long-term monitoring and large number of
epochs the selected sample is very pure, generally free from
contaminating stars of other types often mimicking Cepheids. Its
completeness is high at 90\% level for classical Cepheids -- tested
using recent samples of Galactic Cepheids: ASAS-SN, ATLAS, Gaia DR2 and
Wise catalog of variable stars. Our comparisons indicate that the
completeness of the two latter datasets, Gaia DR2 and Wise catalog, is
very low, at $<\!10\%$ level in the magnitude range of the OGLE GVS
survey ($10.8<I<19.5$~mag). Both these samples are severely contaminated
by non-Cepheids (the purity is 67\% and 56\%, respectively).  

We also present several interesting objects found in the new OGLE
Collection -- multi-mode pulsators, first Galactic candidates for
eclipsing systems containing Cepheid, a binary Cepheid candidate.

New OGLE Collection of Galactic Cepheids is available for the
astronomical community from the OGLE Internet Archive in similar form as
previous parts of the OGLE Collection of Variable Stars.}
{Stars: variables: Cepheids -- Stars: oscillations  -- Galaxy: center --
Galaxy: disk -- Catalogs}

\Section{Introduction}
Cepheid variables belong to the most important tools of modern
astrophysics. These are pulsating giants and supergiants with periods from
a fraction of a day to over 100~d. Their physical parameters, in
particular the luminosity, are well-correlated with the pulsating period
what makes these stars a very accurate tool for the distance
determination. High luminosity of Cepheids allows one to measure precise
distances in the local Universe up to several tens of Mpc. These stars are
also very good empirical benchmarks of the stellar pulsation and evolution
theories.

The term Cepheids does not describe a homogeneous group of pulsating
stars. In fact, there are three groups of objects within this category
with completely distinct properties and evolutionary status, namely
classical Cepheids (CEP), type~II Cepheids (T2CEP) and anomalous Cepheids
(ACEP).

Classical Cepheids are young ($\approx10{-}400$~Myr), Population~I giants
and supergiants evolving through the pulsation instability strip in the
Hertzsprung--Russell diagram. Their pulsation periods are correlated not
only with the luminosity but also with age (Bono \etal 2005, Anderson
\etal 2016) making these stars a potential tool for studying the past of
the systems in which they live in.

On the other hand, type~II Cepheids are much older and less massive
Population~II giants which at certain evolutionary phases can cross the
instability strip and reveal pulsations similar to those of classical
Cepheids. Finally, the anomalous Cepheids are the most enigmatic group --
pulsating giants less luminous than classical Cepheids showing Cepheid-like
light curves. Their evolutionary status has not been firmly established
yet. It is believed that pulsations in these stars are a result of
evolution of either single low-mass, low-metallicity stars or a remnant of
coalescence of a low-mass binary system.

Although the first Cepheids, $\eta$~Aql and $\delta$~Cep, were discovered
back in 18th century by Edward Pigott and John Goodricke, respectively, the
scientific career of these stars started blooming more than a century ago
when Henrietta Leavitt discovered multitude of such objects in the Small
Magellanic Cloud (Leavitt 1908) and subsequently the famous
Period-Luminosity relation was established (Leavitt and Pickering 1912).

Since then the number of known Cepheids gradually increased. The
breakthrough occurred in the 1990s when variable stars became a precious
by-product of the first large-scale sky surveys concentrating on the
detection of gravitational microlensing events (MACHO -- Alcock \etal 1995,
EROS -- Beaulieu \etal 1995, OGLE -- Udalski \etal 1992). Long-term
photometry of millions of stars allowed the detection of large samples of
variables of all known classes, as well as new types of stellar variability
(Pietrukowicz \etal 2017).

The Optical Gravitational Lensing Experiment (OGLE -- Udalski, Szymañski
and Szymañski 2015a) has played the leading role in the variable stars
field since its start in 1992. During all phases of the OGLE survey, big
samples of different types of variables were discovered, carefully
classified, and then released to the astronomical community. They become
the basis for a large number of further studies.

The OGLE Collection of Variable Stars (OCVS, Soszyñski \etal 2018 and
references therein) encapsulates various OGLE contributions, catalogs and
data in the field of variable stars and is the largest collection of
classified variables in modern astrophysics. Currently, it consists of
almost a million of well characterized genuine periodic variable objects of
different types.

OCVS contains large samples of Cepheids discovered by OGLE. The sample of
Cepheids from the Magellanic System (the Large and Small Magellanic Clouds
and the Magellanic Bridge) counts over 10\,000 objects (Soszyñski \etal
2017a, 2018). Such a rich collection allowed detailed studies of the
structure of the Magellanic System in young population
(Jacyszyn-Dobrzeniecka \etal 2016, Inno \etal 2016) and provided many
unique objects like Cepheids in eclipsing systems (Udalski \etal 2015b) or
multi-mode pulsators (Soszyñski \etal 2015b, Smolec \etal 2018).

OCVS also contains Cepheids from the Galaxy. Over 1000 Cepheids of all
classes were found in the OGLE fields in the Galactic bulge (Soszyñski
\etal 2017b). A small sample of Cepheids was also detected in the OGLE
pilot study of the Galactic disk fields (Pietrukowicz \etal 2013).

Only about 900 Galactic classical Cepheids has been discovered so far
(Pietru\-kowicz \etal 2013). The vast majority of them are bright nearby
objects located closer than 4~kpc from the Sun. Thus, this sample is of
little use for studying the general structure of the Milky Way -- the basic
application of classical Cepheids in the Galaxy. Keeping this in mind, the
OGLE survey has launched in 2013 a new large-scale program -- The Galaxy
Variability Survey (GVS) -- aiming at the variability detection and census
of objects in the large sky area of over 3000 square degrees around the
Galactic plane seen from the OGLE observing site (Las Campanas Observatory,
Chile) and the outer Galactic bulge (Udalski \etal 2015a).

Although the GVS survey has not been completed yet, large areas of the
observed sky are ready for exploration and the preliminary results have
already been reported (Udalski 2017). Here we present the details of the
OGLE search for Cepheids in the GVS fields. This is our first attempt to
extract these important stars from over a billion of regularly observed
objects in the GVS fields.

After a very careful analysis of the stability of multi-season light curves
of the Cepheid candidates we finally classified 1339 objects as classical
Cepheids, 316 as type~II Cepheids and 25 as anomalous Cepheids. 1167
classical Cepheids presented here are new discoveries. Together with
earlier OGLE Cepheid discoveries (Pietrukowicz \etal 2013, H\"ummerich
and Bernhard 2013, Soszyñski \etal 2017b) the OGLE Galactic Cepheid
Collection counts 1428 classical Cepheids. The number of type~II Cepheids
is 1240 (after including 924 objects from Soszyñski \etal 2017b) and
anomalous Cepheids -- 53 (with 28 objects detected earlier: Soszyñski
\etal 2017ab). The total number of Galactic Cepheids in the OGLE Collection
is currently 2721.

The new OGLE Collection more than doubled the number of known classical
Cepheids in our Galaxy which reaches now 2476. What is more important, the
photometric range of the GVS: $10.8<I<19.5$~mag allows for the first time a
very detailed study of the Milky Way structure as seen in the young stellar
population -- up to the Galactic disk boundary (Skowron \etal 2018, Mróz
\etal 2018).

The sample of Cepheids presented in this paper is limited to stars brighter
than $I\approx18$~mag. Also, our search for Cepheids in some of the fields
is preliminary, because the number of collected epochs was below our
standard limit of at least 100 epochs. Thus, we expect that there might be
a number of missed Cepheids, in particular in the regions close to the
Galactic plane that are the most obscured by the dust. We plan to update
the OGLE sample of the Galactic Cepheids in the future when the next
generation search is completed and observations of additional GVS fields
are finished.

\Section{Observations}
OGLE observations of Galactic Cepheids come from a large-scale survey,
GVS, conducted since 2013 as one of the sub-surveys of the OGLE-IV
phase. Photometric observations obtained with the 1.3-m Warsaw telescope
equipped with the 32-CCD detector mosaic camera, located at Las Campanas
Observatory, Chile (Observatory is operated by the Carnegie Institution
for Science), started on January 17, 2013 and continue up to now. GVS
consists of three parts: Galactic disk fields extending in Galactic
longitude: $190\arcd<l<345\arcd$ (GD fields), Galactic fields of
$20\arcd<l<60\arcd$ (DG fields) and the outer Galactic bulge defined as
an area roughly: $-15\arcd<l<20\arcd$, $-15\arcd<b<15\arcd$ (BLG fields)
-- see Fig.~1. (RA,DEC) and ($l,~b$) coordinates of all OGLE pointings
in these sections of the sky can be found and downloaded from the OGLE
Web page ({\it http://ogle.astrouw.edu.pl} {\sf Sky Coverage
$\rightarrow$ OGLE-IV Fields} tabs).

The central part of the Galactic bulge have been regularly monitored by
OGLE since 1992 for gravitational microlensing phenomena. The results of
the OGLE extensive search for Cepheids in the microlensing BLG fields were
presented in Soszyñski \etal (2017b). More than 1000 Cepheids of all types
were discovered in these fields. In this paper we present results of our
search for Cepheids in additional part of the BLG fields -- those
complementing the strip $-5\arcd<b<5\arcd$ around the Galactic plane. These
limits are a direct consequence that our priority was the detection of
classical Cepheids which rarely lie outside such a strip. Although not all
from these fields have reached the typical number of epochs requested by
OGLE for our variability search ($>\!100$ observations), the vast majority
of fields were close to that limit.

Observations of the OGLE GD fields reached about 100 epochs after 2016
season and the preliminary results for the OGLE search for Cepheids and
RR~Lyr-type stars were presented in Udalski (2017). Up to now about 150
observations of these
\begin{landscape}
\begin{figure}[htb]
\centerline{\includegraphics[width=20cm]{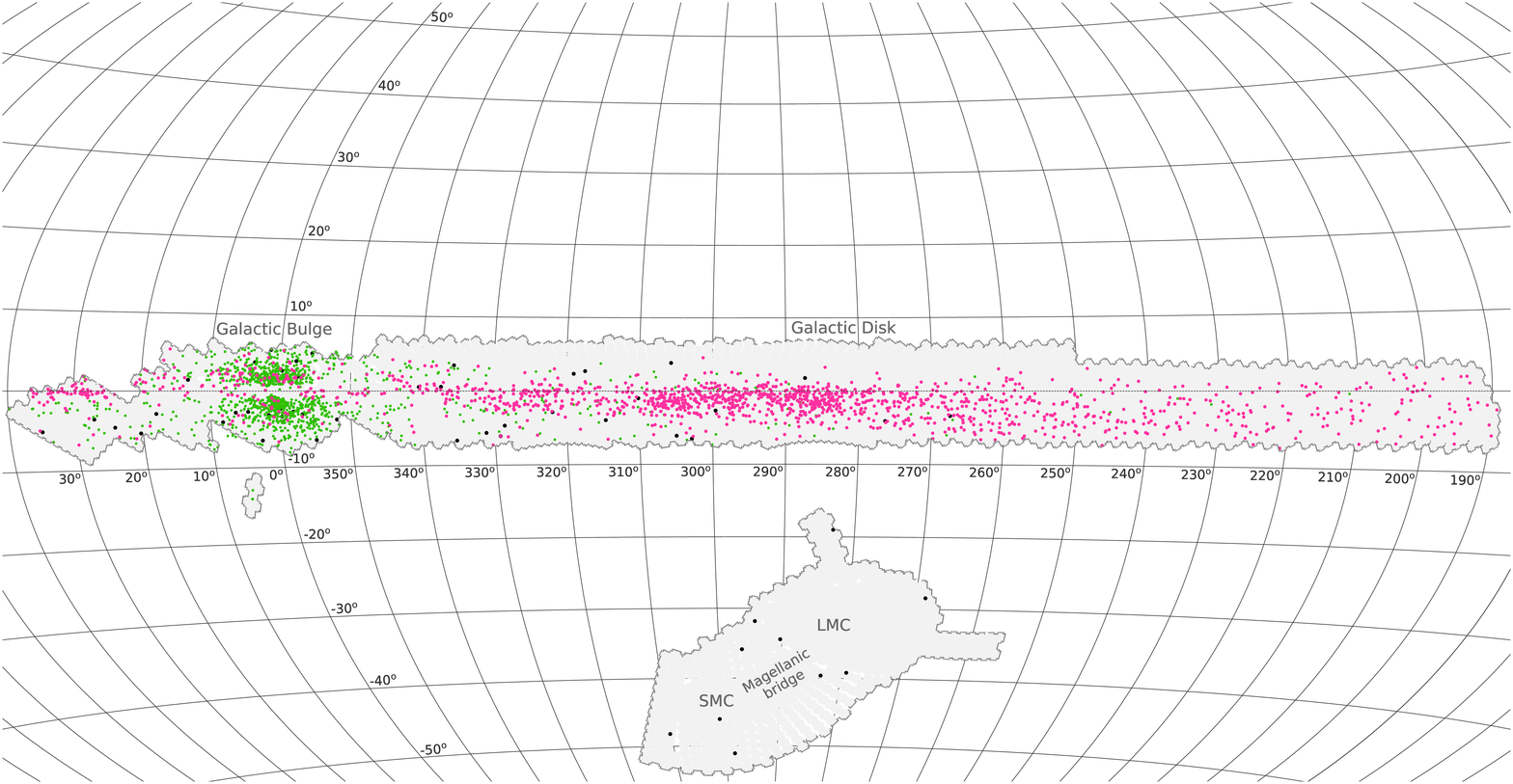}}
\vskip 12pt
\FigCap{Cepheids from the OGLE Collection of Variable Stars in the sky.
Classical Cepheids -- magenta dots, type II Cepheids -- green dots,
anomalous Cepheids -- black dots.}
\end{figure}
\end{landscape}

\noindent
fields spanning over five years were collected. Long span turned out to be
crucial in our search for Cepheids to exclude large population of stars
mimicking Cepheid light curves. In 2017 we extended coverage of the OGLE GD
fields by adding pointings covering area $3\arcd<|b|<6\arcd$ in a large
range of Galactic longitudes of GD fields. Part of these GD extension
fields has also been analyzed here.

The number of observations of the OGLE DG fields has reached the OGLE
requirements after 2017 observing season. Thus, these fields were also
included in our search for Galactic Cepheids.

All observations carried out in the GVS are relatively shallow with the
exposure time of 25~s and 30~s in the {\it I} and {\it V}-band,
respectively. During worse weather conditions the exposure times were
sometimes extended to keep the signal at similar level. The majority of
observations were taken in the {\it I}-band for variability studies and
about 10 epochs were secured in the {\it V}-band for color information.
The range of reliable photometry in the GVS is about $10.8<I<19.5$~mag.

OGLE {\it VI}-band photometry has been calibrated to the standard
Johnson-Cousins system using observations of thousands secondary standards
from the already calibrated OGLE-IV fields observed during the same nights
as the target field. The procedures were similar to those described in
Udalski \etal (2015a). The accuracy of the photometric zero points of the
OGLE Galactic field databases is about 0.02~mag.

Astrometry of the OGLE Galactic fields was carried out in a similar way as
for other OGLE-IV targets (Udalski \etal 2015a). 2MASS coordinate system
(Skrutskie \etal 2006) was the base of the OGLE field astrometric
solutions. The accuracy of transformations between the frame pixel grid
(X,Y) and RA/DEC coordinates is about 0\zdot\arcs1--0\zdot\arcs3 and the
systematic errors of the 2MASS grid are of the order of 0\zdot\arcs1.

\Section{Search for Periodic Variable Stars}
Because the number of photometric data in the observed GVS fields still
increases, we decided to carry out the first search for periodic variable
stars on preselected sub-sample of stars. The natural and simplest
criterion for preselection of a variable candidate is its detection by the
OGLE photometric pipeline (Udalski 2003) on the difference images. The
stars which leave measurable residua on difference images are certainly
variable, thus they should be included in further analysis. While such a
threshold selects true variable objects it also rejects fainter stars for
which reasonable photometry can still be derived and variability, if large
enough, can be detected.  The typical limit of our preselection is
$I\approx18.0$~mag.

Such a preselection does not basically affect Galactic Cepheids, although a
large number of fainter RR~Lyr stars may be missing.  All objects that
are detected in the OGLE GVS (altogether about 1.5 billion) will be
searched for variability in the next generation final analysis.

In the case of the GD fields the requested number of detections on the
difference images was set to ten. In the GD extended and DG fields we
required at least seven detections and in the BLG fields 5--7 detections
depending on the number of collected epochs.

After the preselection, 1\,794\,420 variable candidates were left for
further analysis. In the next step all these objects were subject of the
period analysis. We carried out the period search using two algorithms
-- Fourier based analysis -- {\sc Fnpeaks} by
Z.~Ko³aczkowski\footnote{\it
http://helas.astro.uni.wroc.pl/deliverables.php?lang=en\&active=fnpeaks},
and Analysis of Variance (Schwarzenberg-Czerny 1989). We decided to use
both methods because of relatively limited number of epochs  and similar
sidereal time of observations (so that the dataset was prone to
aliasing). In the case of discrepant results from both algorithms the
real periodicity was selected after careful visual check of the variable
candidate.  

To filter-out non-Cepheid candidates the Fourier series were fitted to
light curves of the periodic variable candidates which passed the previous
step. As usual in the OCVS we fitted cosine Fourier series up to the fifth
order and calculated the amplitude ratios $R_{21}$, $R_{31}$ and phase
shifts $\phi_{21}$, $\phi_{31}$ (Simon and Lee 1981). For preselection of
Cepheid candidates we defined large areas on $R_{21}$ \vs $\log P$ and
$\phi_{21}$ \vs $\log P$ surfaces additionally limiting $\log P$:
\vspace*{-5pt}
$$-0.58<\log P<1.7,$$
\vspace*{-23pt}
$$0.03<R_{21}<0.8,$$
\vspace*{-17pt}
$$\phi_{21}< 2.5\quad{\rm for}\quad 0.9~{\rm d}<\log P<1.2~{\rm d}.$$

These regions embed with large margins the area occupied by Cepheids and
other important pulsating stars -- RR~Lyr -- in the diagrams based on the
OGLE Magellanic Cloud samples containing thousands of these objects (\eg
Soszyñski \etal 2008a -- Fig.~5).

In the GVS regions located close to the Galactic bulge, namely OGLE DG and
BLG (shallow) fields, where the RR~Lyr stars significantly outnumber
short period Cepheids, we decided to put stronger constraint on the range
of searched periods:
\vspace*{-5pt}
$$0.0<\log P<1.7.$$

Short period Cepheids can be easily misclassified with RR~Lyr stars.  By
limiting our search to $P>1.0$~d we get rid of practically all RR~Lyr
contaminants. Our sample of Galactic Cepheids will be complemented with the
potentially missing short period Cepheids in these areas when the OGLE
search for RR~Lyr is completed.

Objects that passed this cut were than subjected to the first pass visual
inspection to filter out obvious non-pulsating stars. It has been quickly
realized that the number of non-Cepheid variables in the preselected sample
is huge. The main contaminants included eclipsing and ellipsoidal stars,
long period variables (LPVs) and spotted stars which may very well mimic
the pulsating light curves. Preliminary results of our search for Galactic
Cepheids after this step were presented in Udalski (2017).

Precise filtering out of the contaminants in our Cepheid sample was not
trivial. Eclipsing stars can be removed relatively easy, when the number of
collected epochs is large enough and coverage of eclipses is sufficient. On
the other hand, light curves of LPVs and spotted stars are usually unstable
in the time scale of months to years. Thus, it is crucial to have coverage
spanning a few years to filter out these types of non-Cepheids. This is the
reason why we decided to collect one additional season of observations
after the preliminary OGLE Cepheid sample was selected (Udalski 2017).

One of the main selection criterion in the final cleaning of the
preliminary OGLE sample was the stability of the long term light curve. We
did it extremely carefully to avoid removing multi-periodic Cepheids which
also present variable shape of light curves. Each case which resembled a
beat Cepheid was analyzed individually.

The next important selection criterion was position of a Cepheid candidate
on the color--magnitude diagram (CMD). This test is very useful in distant
galaxies like the Magellanic Clouds where all objects are at roughly
similar distance. Cepheids populate well defined region of the instability
strip on the CMD and the selection of candidates is, thus,
straightforward. In the case of the Galactic CMDs the situation is much
more complicated because stars (and Cepheids) are located at different
distances from hundreds of parsecs to kiloparsecs. Additionally, large
interstellar extinction near the Galactic plane can significantly dim and
redden Galactic objects. The typical CMDs of the Galactic disk fields look
like those presented in Szymañski \etal (2010) -- the main characteristic
feature is a clear strip of Galactic main sequence stars slanted to redder
color at fainter magnitudes due to increasing reddening. The second much
weaker feature is a redder parallel sequence formed by red clump giants at
different distances.

By checking the position of a Cepheid candidate on the CMD of the field,
one can easily find if this object is a main sequence star. If so it cannot
be a Cepheid even if the light curve has a pulsating-star shape.
Ellipsoidal stars mimicking overtone Cepheids, eclipsing close binary
systems, blue cataclysmic variables, spotted $\alpha^2$ CVn stars can be
removed using this criterion.

Additional and independent verification of the candidates was carried out
based on Fourier parameters of the Fourier light curve
decomposition. Soszyñski \etal (2017a) showed that this can be a very
useful tool for classification of different classes of Cepheids (CEP, ACEP,
T2CEP) because they populate different sequences on appropriate
diagrams. Such diagrams are also useful tools for rejecting
non-Cepheids. We used the following Fourier parameters: $R_{21}$,
$\phi_{21}$, $R_{31}$, $\phi_{31}$ as a function of $\log P$ for
classification of the Galactic Cepheid candidates. Template diagrams were
constructed based on precise OGLE light curves of thousands of Cepheids
from the Large Magellanic Cloud (Soszyñski \etal 2008a).
   
The sample of OGLE Galactic Cepheids after applying all the above mentioned
selection cuts was finally visually inspected by three very experienced
astronomers (IS, PP, AU). In this final vetting step we rejected several
uncertain candidates and left on the final list only the Cepheids accepted
by the vetting team.

In total the OGLE sample of Galactic Cepheids presented in this paper
contains 2721 objects.
\Section{OGLE Collection of Galactic Cepheids}
The structure of the new part of the OCVS -- the Galactic Cepheids -- is
identical to the former parts, in particular the OGLE Collection of
Cepheids in the Magellanic System (Soszyñski \etal 2015a, 2017a, 2018). The
entire collection of the OGLE Galactic Cepheids is available from the OGLE
WWW Page and OGLE FTP Internet Archive:

\centerline{\it http://ogle.astrouw.edu.pl}
\centerline{\it ftp://ftp.astrouw.edu.pl/ogle/ogle4/OCVS/gd/cep}
\centerline{\it ftp://ftp.astrouw.edu.pl/ogle/ogle4/OCVS/blg/cep}
\centerline{\it ftp://ftp.astrouw.edu.pl/ogle/ogle4/OCVS/gd/t2cep}
\centerline{\it ftp://ftp.astrouw.edu.pl/ogle/ogle4/OCVS/blg/t2cep}
\centerline{\it ftp://ftp.astrouw.edu.pl/ogle/ogle4/OCVS/gal/acep}
\centerline{~}

We continue to use the standard OGLE naming convention for detected
Ce\-pheids: OGLE-GD-CEP-NNNN, OGLE-GD-T2CEP-NNNN for Cepheids located in
the OGLE Galactic disk fields. The first twenty numbers for classical
Cepheids have already been occupied by the OGLE-III Cepheids discovered by
Pietrukowicz \etal (2013) in a pilot survey of the Galactic disk based on
the OGLE-III observations. Not all of them can be confirmed as {\it bona
  fide} Cepheids in the OGLE-IV data (Section~5), nevertheless we keep the
numbering for historical reasons. We start the list of OGLE-IV Galactic
classical Cepheids from '0021'. Although none of the type~II Cepheid
candidates presented by Pietrukowicz \etal (2013) survived after OGLE-IV
verification (Section~5) we keep numbering here as well. Therefore, we
start the OGLE-IV Collection of this class from OGLE-GD-T2CEP-0007.

Although the Galactic bulge is a crucial part of our Galaxy, it is kept as
a separate entity in the OGLE Collection. Therefore, we decided to add all
the new OGLE-IV Cepheid detections in the OGLE outer Galactic bulge fields
(designated with prefix BLG) to the already existing OGLE BLG collections
of classical and type~II Cepheids. The numeration of presented in this
paper Cepheids will start from OGLE-CEP-BLG-101, and OGLE-T2CEP-BLG-0932
for classical, and type~II Cepheids, respectively. We add one more
significant digit for the latter Cepheid class as the original three-digit
field was close to overflow.

In the case of Galactic anomalous Cepheids the first seven OGLE objects
were detected in the Galactic halo in the direction of the Magellanic
Clouds (Soszyñski \etal 2017a). The next twenty in the Galactic bulge
(Soszyñski \etal 2017b) and, very recently, one more in the direction of
the Magellanic Bridge. All of them have designations OGLE-GAL-ACEP-NNN. For
compatibility, we add all new detections of anomalous Cepheids in both: the
OGLE-IV Galactic disk (GD, DG) and outer Galactic bulge (BLG) fields to
this list, starting from OGLE-GAL-ACEP-029.

The data provided in the OCVS for OGLE-IV Galactic Cepheids include
astrometry (RA/DEC J2000.0), mean magnitudes in the {\it I} and {\it
V}-bands, pulsation period, epoch of the maximum of light, {\it I}-band
amplitude and Fourier parameters from the Fourier light curve
decomposition. After rejecting obvious outlying points, the pulsation
periods have been tuned up using the {\sc Tatry} program
(Schwarzenberg-Czerny 1996). For details see {\sc Readme} file in the
appropriate sub-archives of the OCVS.

In total the search presented in this paper provided 1680 objects: 1339
classical Cepheids, 316 type II Cepheids (154 BL~Her type, 143 W~Vir type,
including four peculiar W~Vir type, Soszyñski \etal 2008b, and 19 RV~Tau
type) and 25 anomalous Cepheids. The separation in pulsation period between
different groups of type II Cepheids was the same as in other parts of the
OGLE Collection -- shorter than 4~d for BL~Her, shorter than 20~d for W~Vir
Cepheids.

Fig.~1 presents the sky map in the Galactic coordinates with the positions
of the OGLE Cepheids: CEP, T2CEP and ACEP marked with different colors.
\Section{Completeness of the OGLE Cepheid Sample}
OGLE Collection of Galactic Cepheids supplemented with OGLE microlensing
fields detections (Soszyñski \etal 2017b) contains 2721 stars -- classical,
type~II and anomalous Cepheids. Because the vast majority of Galactic
classical Cepheids are bright objects with $I<18$~mag, the completeness of
the OGLE Collection for this type of pulsators is high. We expect that our
preselection of objects which was the first step of our search for Cepheids
only marginally affects completeness of the OGLE sample (only the very
highly reddened objects near the Galactic plane and Galactic center may be
obscured so much that they are below our preselection threshold; a
$P=3$~d Cepheid at $d\approx20$~kpc from the Sun should be detectable even
at the extinction of $A_I\approx4$~mag). Due to technical reasons, the OGLE
footprint of the Galactic fields in the sky contains non-covered regions
reaching 5--7\% of the area (gaps between CCD detectors of the OGLE-IV
camera). Thus, even the full completeness of the OGLE observed fields means
$\approx94\%$ absolute completeness.

In the case of classical Cepheids we estimate that the completeness of the
OGLE Collection is at $\approx90\%$ level. For type~II Cepheids of BL~Her
and W~Vir type the completeness is still high -- about 80--90\%. We note
here that only single RV~Tau type stars are included to the OGLE sample at
this moment because presented here search has not been focused on such
objects. In the case of anomalous Cepheids the completeness is difficult to
assess. Generally these stars are fainter and harder to distinguish from
other Cepheid types and RR~Lyr stars (\cf Soszyñski \etal 2017b). Thus, our
completeness may be lower.

High completeness of the OGLE Collection of Galactic Cepheids can be
verified by comparison with other Cepheid detections from the
literature. On the other hand, we can assess purity of other surveys which
have recently claimed the discovery of significant number of Galactic
Cepheids. In this latter case we may compare Cepheid candidates with all
objects in the full range of the OGLE-IV databases (even the faintest --
down to $I\approx19.5$~mag) as the position of each candidate is
known. However, when assessing the OGLE search completeness we only use
those stars which are currently in the OGLE Collection and which should
pass our preselection cut (thus, be brighter than $I\approx18$~mag).

For years there have been two major sources of Galactic Cepheids in the
literature: the General Catalogue of Variable Stars (GCVS, Artyukhina \etal
1995) and the All Sky Automated Survey (ASAS) catalog (Pojmañski
2002). Unfortunately, the majority of Cepheids listed in these sources are
nearby objects ($d<4$~kpc) -- bright and mostly saturated in the OGLE GVS
observations. The original GCVS sample is also non-homogeneous and,
sometimes, hard to verify because objects listed there came often from old
observations obtained with out-of-date techniques.

After presentation of the preliminary sample of the OGLE Galactic
Cepheid Collection (Udalski 2017) a few new datasets of variable stars
containing Galactic Cepheid candidates have been published. Below, we
compare the content of all these datasets with the OGLE Collection of
Galactic Cepheids. 

\subsection{GCVS and OGLE Collection of Galactic Cepheids}
For comparisons with the GCVS we have used objects included in the list of
Galactic classical Cepheids prepared by Pietrukowicz \etal (2013,
Section~6). This list is based on the original GCVS list (Artyukhina \etal
1995) with upgrades in the following years (Samus \etal 2017). Objects
from GCVS have been additionally verified and often reclassified \vs the
original GCVS classification based on modern data and AAVSO VSX Catalog.

Current version of the Pietrukowicz \etal (2013) list contains 713
classical Cepheids defined as GCVS objects (they have GCVS as main ID).

328 of these classical GCVS Cepheids are located in the OGLE sky coverage
footprint. As expected the majority of them (224) are saturated in OGLE
images. 12 fall in uncovered part of the OGLE area. 90 classical GCVS
Cepheids were detected in the OGLE data and confirmed as genuine
Cepheids. One could not be verified because it is located very close to a
bright saturated star.

\subsection{ASAS Survey and OGLE Collection of Galactic Cepheids}
The ASAS survey has detected and classified a large number of bright
variables considerably increasing the number of known bright Cepheids.
This survey collected large number of epochs, thus, the classification of
variables is rather sound and homogeneous. Unfortunately, the ASAS and OGLE
surveys overlap only in a very limited magnitude range -- faint end of ASAS
and bright end of the OGLE GVS. Similarly to the GCVS case -- many ASAS
detections are saturated in OGLE images.

105 ASAS classical Cepheids are listed by Pietrukowicz \etal (2013). 60 are
located in the OGLE footprint. 22 were detected and confirmed by OGLE. The
remaining 36 are saturated and two fell into non-observed area of the
fields.

\subsection{OGLE-III Pilot Survey and OGLE Collection of Galactic Cepheids}
The most convenient and comparable in quality dataset of Cepheids suitable
for checking the completeness of the present search is the OGLE Galactic
disk pilot study by Pietrukowicz \etal (2013). Unfortunately, this survey
focused on the OGLE-III transiting planet search, covered only 7.12 square
degrees. On the other hand, the pointings were in the stellar rich Galactic
fields and the exposures were much longer than those of the OGLE-IV
GVS. Pietrukowicz \etal (2013) found 20 classical Cepheids and six
candidates for type~II Cepheids in these OGLE-III fields.

All classical Cepheids, except one, OGLE-GD-CEP-0020, which fell in
the uncovered part of the field, were independently detected in the OGLE-IV
GVS. However, OGLE-GD-CEP-0013 turned out to have completely different
shape of the light curve compared with a decade earlier observations during
OGLE-III phase. Variable shape of the light curve indicates that this is
rather a spotted star than a Cepheid. Therefore OGLE-GD-CEP-0013 has been
retracted from the list of OGLE Cepheids. The remaining 18 OGLE-III
classical Cepheids have been confirmed implying high, $>\!90\%$
completeness of the OGLE-IV Galactic Cepheid Collection.

On the other hand, the light curves of all six candidates for type~II
Cepheids show in the OGLE-IV GVS data different and variable shapes compared
to the original OGLE-III ones (Pietrukowicz \etal 2013). This clearly
indicates that these objects are also rather spotted stars than genuine
type~II Cepheids. Therefore, we also rejected them from the OGLE-IV
Collection.

\subsection{ASAS-SN Survey and OGLE Collection of Galactic Cepheids}
ASAS-SN project has released a variable star catalog based on its all-sky
photometry (Jayasinghe \etal 2018). The range of this search for variables
is relatively shallow ($V<17$~mag). However, the number of collected epochs
is large which makes classification significantly easier. Classification is
obtained using automatic machine learning algorithms.

ASAS-SN catalog contains 315 objects classified as classical or type~II
Ce\-pheids. 66 stars are located in the footprint of the OGLE-IV GVS.
However, two are overexposed in the OGLE images, two are located in the
gaps between the OGLE subfields, six have too few observing points to be
analyzed in this OGLE search (they are located close to the field edge).
The remaining 56 ASAS-SN objects can be verified using OGLE photometry.
Five are misclassified by ASAS-SN (non-Cepheids) and one (ASAS-SN
J081259.7-442549) was overlooked by OGLE due to bad period value (1-d alias
with resulting $P<1$~d, thus, not included here). The remaining 50 common
stars are genuine Cepheids. This comparison of the OGLE Collection
indicates its high completeness in the range covered by ASAS-SN.

\subsection{ATLAS Survey and OGLE Collection of Galactic Cepheids}
The Asteroid Terrestrial-impact Last Alert System survey (ATLAS, Tonry
\etal 2018) designed for finding near-Earth asteroids (NEAs) has collected
during the past years a large set of images of the sky north of declination
$\delta=-30\arcd$ suitable also for analysis of stellar variability. In
April 2018 a huge catalog containing about 450\,000 of variable object
candidates based on photometry collected during the ATLAS survey was
published. Additionally, the whole set of individual measurements
containing over 100 epochs for each individual object was released (Heinze
\etal 2018).

Because the automatic classification approach used by the authors of the
ATLAS catalog does not meet the OGLE strict criteria of variability
analysis we decided to independently classify the ATLAS Cepheid
candidates. We downloaded light curves of all stars preliminary designed as
pulsating (all types) in the ATLAS catalog and many more objects where
Cepheid candidates could be hidden. Altogether we analyzed over 3000 ATLAS
light curves, independently in two filters used. They were run through the
OGLE data pipeline -- identical as used for objects from the OGLE
Collection. Because we were mostly interested in classical Cepheids for the
Galactic structure studies we concentrated on selection of only this type
of ATLAS Cepheids neglecting other classes.

438 classical Cepheids were found during our search of the ATLAS data. We
compared the list of these objects with Cepheids from the OGLE Collection
and lists of other known classical Cepheids. 170 objects are the new
discoveries -- Cepheids undetected by previous searches.

101 out of 438 ATLAS classical Cepheids are located in the OGLE
footprint. 89 could be verified using OGLE databases (the remaining objects
are either overexposed in OGLE images or fall into gaps between OGLE camera
detectors). 87 Cepheids were already found by OGLE during the search
presented in this paper. Two missing objects are certainly classical
Cepheids but the number of collected epochs by OGLE was too low for passing
preselection and period determination steps. Comparison of ATLAS data and
OGLE collection indicates again that the completeness of the OGLE
collection is almost perfect.

\subsection{Gaia DR2 Cepheids and OGLE Collection of Galactic Cepheids}
At the end of April 2018 the Gaia satellite team made available its second
data release, DR2, including, among others a set of variable stars detected
and characterized by the Gaia pipeline (Holl \etal 2018). Gaia DR2 includes a
sample of Cepheids of all types (classical, type~II and anomalous) and the
main, most reliable set of Gaia Cepheid candidates -- SOS Cep\&RRL
(Clementini \etal 2018) -- contains 9575 stars claimed as Cepheids. The vast
majority of them are located in the Magellanic System. Virtually all of them
have already been included in the OGLE Collection of this region of the sky
(see the comparison in Udalski \etal 2016 and Soszyñski \etal 2018). Large
number of OGLE Cepheids were used by Holl \etal (2018) to train the Gaia DR2
automatic classifiers.

The remaining part of the Gaia DR2 Cepheid candidates is located all over
the sky. The photometric ranges of the Gaia photometry and the OGLE survey
are comparable. Thus, the huge OGLE footprint of the GVS fields presented in
this paper ($\approx1800$ square degrees) enables direct comparison of the
datasets and verification of the quality of the Gaia DR2 Cepheid sample in
the sky regions where the vast majority of Galactic classical Cepheids
reside (Galactic disk). On the other hand, the completeness of the
presented here collection of OGLE Galactic Cepheids can be checked based on
independent Gaia DR2 set.

559 Gaia objects from the DR2 SOS Cepheid set are located in the footprint
of the OGLE Galactic fields observed so far, \ie the central Galactic bulge
fields observed for microlensing and already searched for Cepheids
(Soszyñski \etal 2017b) as well as Galactic disk and outer Galactic bulge
fields presented in this paper. Not all of them can be verified as Gaia
also observes brighter stars which are saturated in the OGLE
images. Altogether 344 Gaia objects can be cross-identified with stars in
the OGLE databases and verified with the OGLE most recent photometry. The
remaining Gaia objects are either too bright for the OGLE camera (thus
saturated) or fall in the unobserved gaps between the OGLE mosaic camera
CCD detectors.

Out of these 344 Gaia objects 231 can be confirmed by OGLE photometry as
{\it bona fide} Cepheids. The purity of the Gaia DR2 Cepheid sample in the
Galactic disk region of the sky is, then, about 67\%. The remaining Gaia
objects which are certainly non-Cepheids are in the vast majority eclipsing
stars (32\%) and spotted stars (22\%) but also LPVs and sometimes even
non-variable objects.

The OGLE search presented in this paper yielded 223 Cepheids out of 231
Gaia DR2 verified ones, \ie 97\%. The missing objects are located usually
at the edge of detectors. Thus, the number of their observations at the
moment when our search for Cepheids started was too small either to pass
the preselection criterion or to derive the correct period. In the case of
one missing object (GAIA 5521400228203695232) the period was so close to
1~d that finding this double mode Cepheid from the ground would be a
miracle.

The total number of all types Cepheids in the OGLE Collection (anomalous,
classical, type~II but excluding RV~Tau which have not been carefully
searched by OGLE yet) is 2538 (45 anomalous -- not counting eight Galactic
ACEPs located in the direction of the Magellanic System, 1428 classical,
1065 type~II -- BL~Her and W~Vir). They are fainter than $I\approx
10.8$~mag, \ie in the most interesting magnitude range for many new
applications and discoveries.

Provided statistics indicate that the completeness of the Gaia DR2 Cepheid
sample \vs the OGLE Collection in the Galactic disk and bulge sky regions
is very low -- only at the $231/2538=9.1\%$ level. In part this may be
caused by low number of epochs collected by Gaia up to the DR2 dataset and
the high stellar density of the fields. These statistics may improve with
further Gaia data releases.

\subsection{WISE Catalog of Variables and OGLE Collection of Galactic Cepheids}
Another relatively large catalog of variable stars containing the Galactic
Ce\-pheid candidates was published in July 2018 by Chen \etal (2018). This
catalog is based on all sky mid-infrared photometry collected during the
WISE satellite mission. The authors analyze and classify the infrared light
curves of individual objects. This catalog includes 1312 objects classified
as Cepheids and a number of Cepheid-like objects.

We downloaded the data for Cepheid and Cepheid-like classifications from
the WISE catalog (Chen \etal 2018). The sample consists of 1772
objects. 509 of them are located in the OGLE footprint. It is well known
that the classification of pulsating stars is difficult in the infrared
bands and often leads to misclassifications. Therefore, we have first
verified purity of the WISE catalog. We cross-identified WISE Cepheid
candidates with objects from the OGLE databases.

434 Cepheid candidates from the WISE catalog can be verified in our optical
OGLE databases. The remaining objects are overexposed in the OGLE images,
are located in the gaps between OGLE camera CCDs or are too faint
($I>20$~mag) in the optical range for verification -- likely due to high
extinction. Moreover, in many cases the periods listed in the WISE catalog
are wrong up to several percent (in 25\% cases of the real Cepheids). Thus,
we have calculated proper periods for each of the candidates based on the
OGLE optical data. Wrong periods may be a result of specific pattern of
WISE satellite observations.

244 WISE objects out of 434 having OGLE counterparts can be verified as
genuine Cepheids. Thus, the purity of the sample is at the $244/434=56\%$
level, \ie every second object is not a Cepheid. The largest group of
misclassified objects are eclipsing variables (32\% of the sample).  Some
objects are LPVs or simply non-variables.

During the OGLE search presented in this paper 231 Cepheids from the WISE
verified list were found. Five were missed because of too small number of
epochs or wrong period determination and eight uncertain Cepheid candidates
have not passed the OGLE preselection procedure because they are too faint
($>18$)~mag due to high interstellar extinction close to the Galactic plane
(while being detectable in the infrared). Thus, the completeness of the
OGLE collection estimated with the verified WISE Cepheids remains at the
$231/236=97.9\%$ level. The eight missing faint Cepheid candidates on the
whole OGLE footprint indicate that the number of faint Cepheids overlooked
in our present search should be less than $\approx5\%$.

Because the WISE catalog is all sky collection of variables and in the
infrared the majority of Galactic Cepheids should be detectable (\cf
Skowron \etal 2018) we may estimate the completeness of the WISE
catalog. As already mentioned (Section~5.6) the number of all types of
Cepheids in the OGLE Collection footprint is 2538. Thus, including the WISE
faint Cepheid candidates, the completeness of the WISE catalog is at the
$244/(2538+10)=9.6\%$ level -- again very low compared to the OGLE
Collection.
\subsection{Ground Infrared Surveys and OGLE Collection of Galactic Cepheids}
Our analysis of the detections of infrared Cepheids based on the comparison
of the WISE and OGLE data confirms that the solely infrared detections can
be significantly contaminated by other variables. On the other hand, using
the infrared bands is tempting as one can probe the Galactic plane and
Galactic center regions -- not available for optical bands. Matsunaga \etal
(2016) and D\'ek\'any \etal (2015) reported the discovery of several tens
of classical Cepheids in the optically obscured regions of the Galactic
bulge, based on the IRSF/Sirius (Nagashima \etal 1999) and VVV (Minniti
\etal 2010) surveys, respectively. We tried to cross-identify those objects
in the OGLE {\it I}-band images. We succeeded only in two cases -- stars
\#6 and \#7 on the Matsunaga \etal (2016) list (\#7~=~VVV-29 on the
D\'ek\'any \etal 2015 list). OGLE optical photometry confirms the Cepheid
status and period of the star \#6 (OGLE-BLG-CEP-121). Its {\it I}-band
magnitude is, however, fainter than our preselection limit, so this
variable has not been found during the present search. On the other hand,
the star \#7 cannot be confirmed as a periodic variable in spite of its
reasonable OGLE photometry. This is another example that the infrared
detections should be treated with care.

Three more infrared Cepheid candidates were detected by Tanioka \etal
(2017) in the Galactic disk close to the Galaxy center. Only one object can
be verified with the OGLE optical photometry -- Lp30A. OGLE data confirm
Cepheid classification and this star is designated as OGLE-GD-CEP-1238 in
the OGLE Collection.

Five additional infrared Galactic disk Cepheids were presented by Inno
\etal (2019). While the two brightest objects (ID-1 and ID-4) are known
Cepheids from the OGLE Collection -- OGLE-GD-CEP-1017 and OGLE-GD-CEP-1021,
the remaining three are very faint in the optical {\it
I}-band. Additionally, one of the candidates (ID-2) -- bright in the
infrared -- is blended with much brighter optical companion, thus
practically non-verificable in the optical band. One of the two remaining
candidates (ID-5) reveals noisy Cepheid-like light curve in the OGLE
data. The last one does not show any trace of periodic optical variability,
although we cannot exclude it is hidden in the observing noise as the star
is very faint in the {\it I}-band.
\subsection{Other Detections and OGLE Collection of Galactic Cepheids}
To complete this Section we note that the discovery of 11 new Cepheids
and 88 Cepheid candidates in the Galactic bulge was reported by Kains
\etal (2019). We were able to identify and verify 95 stars. None of
these objects can be confirmed as a Cepheid based on the OGLE extensive
and precise photometry. Most of them are periodic but rather
spotted, ellipsoidal or eclipsing than pulsating variables.

\subsection{Summary of Comparisons}
Summarizing, all these tests clearly indicate that the sample of
Galactic Ce\-pheids presented in this paper has high completeness,
purity and quality. It significantly increases the number of known
classical and other type Cepheids and opens new possibility for studies
of our Galaxy and better understanding of properties of these very
important standard candles.

Please note that the single Cepheids overlooked during our OGLE search
but within its magnitude range and found during the above verifications
with other datasets have been added to the OGLE Collection of Galactic
Cepheids for completeness. It is also worth noting that the OGLE
photometry of misclassifed Cepheid candidates described in Section~5 can
be found in the OGLE Collection of Galactic Cepheids archive.

\Section{OGLE List of Galactic Classical Cepheids}
Because the classical Cepheids are one of the main tools for studying the
Galactic structure we completed a list of all known and verified with
modern observations Galactic Cepheids, trying to keep it as reliable as
possible. This list was originally presented by Pietrukowicz \etal (2013)
and its electronic version became available from the OGLE {\it ftp} site.
With a flurry of new discoveries of new Galactic Cepheids the major update
of the list and its structure was necessary.

A new version of the list of Galactic classical Cepheids is available from
the OGLE {\it ftp} archive:

\centerline{
{\it ftp://ogle.astrouw.edu.pl/ogle/ogle4/OCVS/allGalCep.listID}}

\vspace*{6pt}
Additionally, we prepared a Web page containing the same information and
some visualizations. It is available from the main OGLE Web site:

\vspace*{6pt}
\centerline{{\it http://ogle.astrouw.edu.pl}}

\vspace*{6pt}
\noindent
Please cite Pietrukowicz \etal (2013) paper when using the list in
publications.

The original list was supplemented with the new OGLE discoveries as well as
confirmed detections from other surveys. To keep the purity as high as
possible we do not include in this list Cepheid candidates which cannot be
verified (except for IR Cepheid candidates in the Galactic center). Stellar
contamination in the Galactic fields is so high that only sound detections
enter the list. Currently, the list counts 2476
objects. Cross-identification with other surveys is also provided.  Updates
of the list will be done regularly.
  
\Section{Basic Statistics of the OGLE Collection}
OGLE Collection of Galactic Cepheids contains now 2721 Cepheids from the
Milky Way. 87 classical, 924 type~II and 28 anomalous Cepheids have been
already reported in Soszyñski \etal (2017ab). The remaining OGLE objects
are reported in detail in this paper for the first time. A part of them
have already been detected earlier. Our Collection will be updated when
observations of additional OGLE-IV Galactic fields are completed.

Having so large sample of 1428 OGLE classical Cepheids from the Galaxy we
may compare basic parameters of Cepheids in different environments.
Virtually all classical Cepheids in the Magellanic System have been found
during the OGLE monitoring (Soszyñski \etal 2017a). Young stars from the
Large Magellanic Cloud (LMC) have on average a half of metallicity of the
Milky Way young stars. The metallicity of the Small Magellanic Cloud (SMC)
young population is by an additional factor of two lower than that of the
LMC objects. Thus, direct comparison of these samples provides empirical
data on how properties of Cepheids depend on the chemical composition of
galaxies they live in.

\begin{figure}[htb]
\centerline{\includegraphics[width=10.1cm, bb=25 50 525 520]{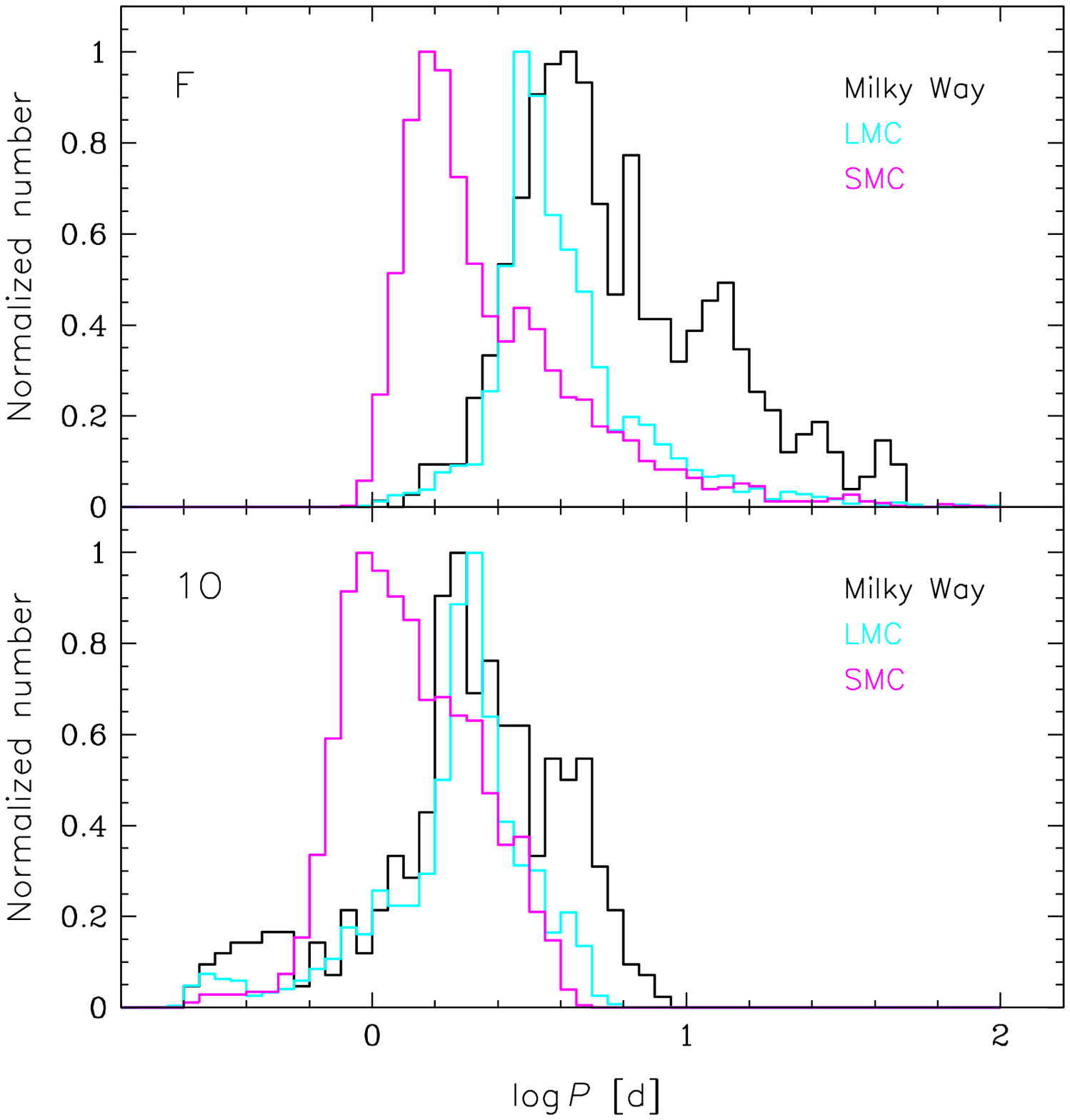}}
\vskip5pt
\FigCap{Distribution of periods of classical Cepheids from the OGLE
Collection of Galactic Cepheids. Similar distributions of periods for
Cepheids from the Magellanic System are shown for comparison: cyan for
the LMC and magenta for the SMC. {\it Upper panel} -- fundamental-mode
Cepheids, {\it lower panel} -- first-overtone Cepheids.}
\end{figure}
Fig.~2 presents distribution of periods for classical Cepheids of
fundamental pulsation mode (upper panel) and the first overtone mode (lower
panel) for stars from the Milky Way and, for comparison, from the
Magellanic System (Soszyñski \etal 2015a). The shift of the pulsation
period distribution in environments of different metallicities is clearly
seen in both pulsation modes. Periods are longer in more metal abundant
environments.

\begin{figure}[b]
\centerline{\includegraphics[width=12.7cm, bb=20 225 575 755]{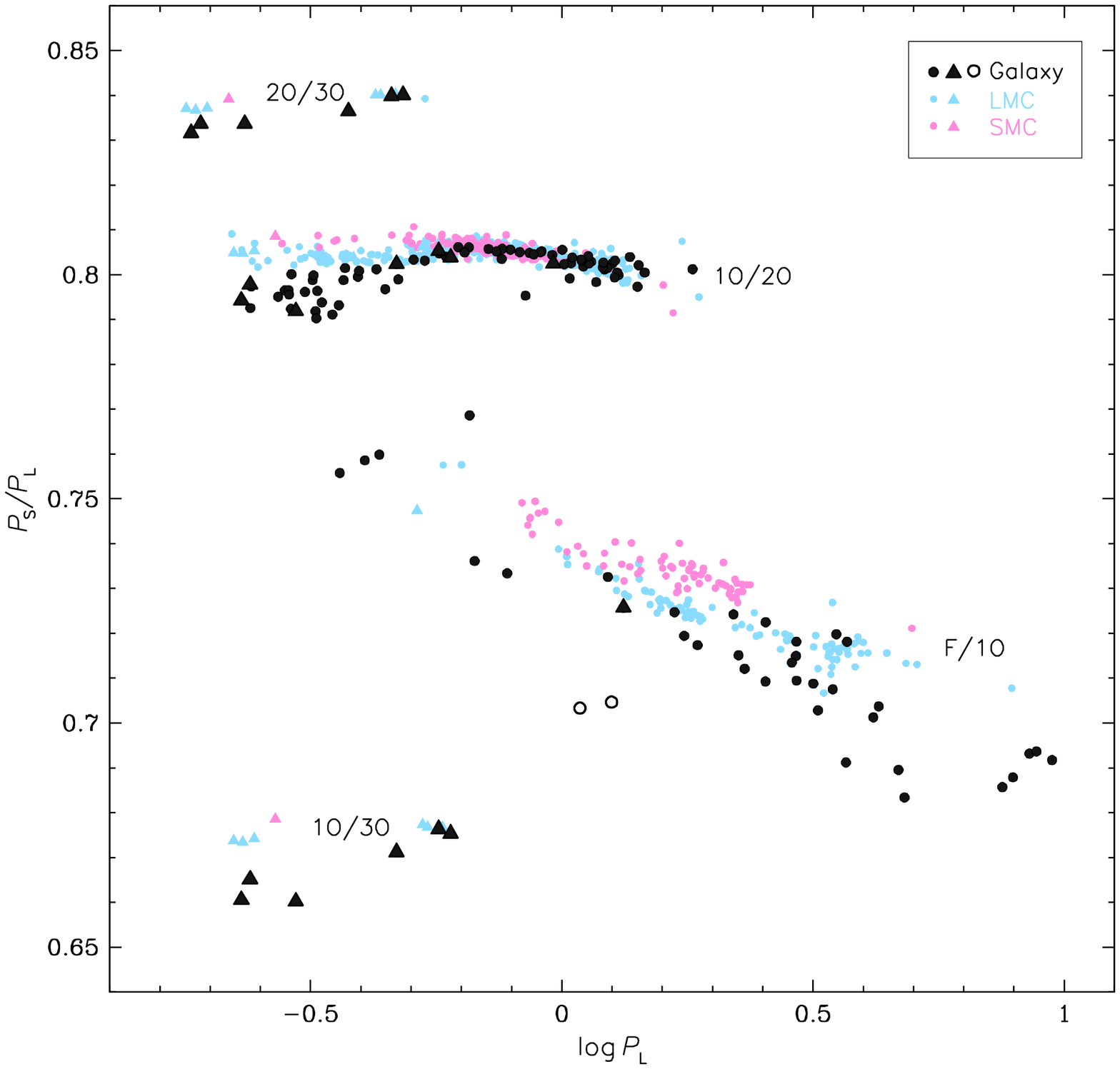}}
\FigCap{Petersen diagram for multi-mode classical Cepheids. Black dots
and triangles show multi-mode Cepheids from the OGLE Collection of
Galactic Cepheids while cyan and magenta signs show multi-mode Cepheids
from the LMC and SMC, respectively. Two open black circles mark position
of two new multi-mode type~II Cepheids.}
\end{figure}
\Section{Multi-periodic Objects}
OGLE Collection of Galactic Cepheids contains many new multi-periodic
objects. The most prominent are Cepheids pulsating in more than one radial
mode. Altogether the OGLE Collection contains 107 multi-mode Cepheids
pulsating either in the fundamental and first overtone or first and second
overtone modes. Six additional objects are triple-mode Cepheids pulsating
simultaneously in the first, second and third overtones. Three of these
unique objects have already been presented in the previous paper (Soszyñski
\etal 2017b). Additional three -- OGLE-GD-CEP-0360, OGLE-GD-CEP-0555 and
OGLE-GD-CEP-0643 -- have been\break found during the GVS. Moreover, we have also
found one triple-mode Cepheid pulsating in the fundamental mode and the
first and second overtones -- OGLE-GD-CEP-1011.
\begin{figure}[b]
\centerline{\includegraphics[width=12cm, bb=30 450 570 770]{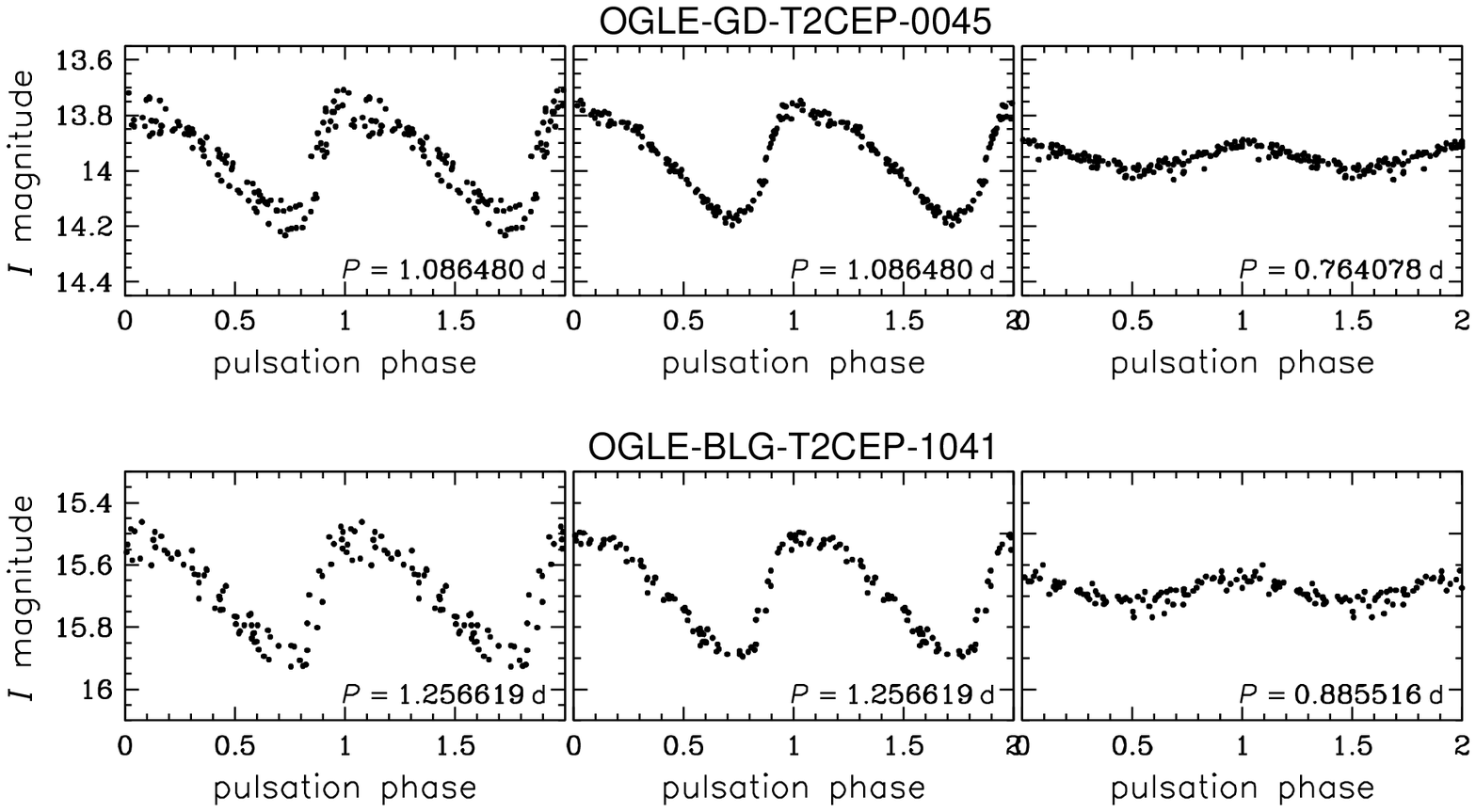}}
\FigCap{ Light curves of OGLE-GD-T2CEP-0045 and OGLE-BLG-T2CEP-1041 --
new candidates for double mode type II Cepheids. {\it Left panel} shows
the original data folded with the fundamental mode period. {\it Middle}
and {\it right panels} present separated pulsations in the fundamental
mode and first overtone, respectively.}
\end{figure}
Fig.~3 shows the Petersen diagram presenting the ratio of the shorter to
longer pulsation periods \vs logarithm of the longer period for the
multi-mode Galactic classical Cepheids from the OGLE Collection (black
dots). For comparison, similar data are plotted for classical Cepheids from
the LMC (cyan dots) and SMC (magenta dots). One can notice that the
characteristic sequences in the Petersen diagram for double-mode stars from
different environments are somewhat shifted one \vs another indicating
again some dependence of pulsation properties on metallicity of Cepheids.

When preparing the diagram we noted that two short period multi-mode
Cephe\-ids originally classified as classical ones are evident outliers
from the sequence of F/1O multi-mode classical Cepheids (Fig.~3). With the
pulsation periods close to one day and the period ratio of 0.70 they
closely resemble the first two multi-periodic type~II Cepheids of BL~Her
type found recently by Smolec \etal (2018). Moreover, their light curves
are almost identical with those prototypes (Fig.~4). Thus, we finally
reclassify these two Cepheids as T2CEP type: OGLE-GD-T2CEP-0045 and
OGLE-BLG-T2CEP-1041 increasing the sample of multi-periodic BL Her stars to
four.

It is very likely that the OGLE Collection contains a sample of multi-mode
Cepheids in which the additional modes are non-radial. Many such objects
have been discovered, for example, in the LMC sample (Soszyñski \etal
2015b, Smolec 2017). However, the search for such objects is out of the
scope of this paper and we leave it for follow-up studies.

\Section{Binary Candidates}
One of the large successes of the OGLE search for Cepheids in the
Magellanic Clouds was the discovery of several Cepheids in binary eclipsing
systems (Soszyñski \etal 2008a, Udalski \etal 2015b). With the
spectroscopic follow-up observations obtained by the Araucaria project, the
first, spectacular measurement of the Cepheid mass with the accuracy of 1\%
was possible (Pietrzyñski \etal 2010). This pioneering measurement provided
an empirical solution of the long-standing discrepancy between the
predicted Cepheid masses by theories of stellar pulsation and evolution --
differing by $>\!\!20\%$.

Since then, several more classical Cepheids in eclipsing systems were
found. Part of them have already been followed-up spectroscopically by the
Araucaria project (Pilecki \etal 2018a). The current status of the OGLE
Cepheids in eclipsing systems can be found in Udalski \etal (2015b).

Our search for Galactic Cepheids yields three very interesting binary
systems. The first one, OGLE-GD-CEP-0069, contains an overtone classical
Cepheid. Its pulsation period is $3.832704\pm0.0001069$~d. It is a member
of a wide eclipsing system with the orbital period of 81.64~d.  Different
width of eclipses may indicate a presence of a disk in the system similar
as in a W~Vir star OGLE-LMC-T2CEP-211 (Pilecki \etal 2018b).

OGLE-GD-CEP-0465, is another eclipsing system containing a classical Cep\-heid
as one of the components, however, pulsating in fundamental mode. The
pulsation period is $6.605564\pm0.000059$~d while the orbital period equals
to 193.83~d. Fig.~5 presents the light curves of OGLE-GD-CEP-0069 and
OGLE-GD-CEP-0465 showing the variability folded with the pulsation period and
eclipsing light curve after removal of pulsations. Both these objects remain
very promising candidates for the first eclipsing Cepheid systems in the
Galaxy. However, their status must be confirmed spectroscopically to exclude
a by-chance coincidence of two independent variables. This is unlikely in the
rather empty stellar fields where these stars reside.
\begin{figure}[htb]
\centerline{\includegraphics[width=12.3cm, bb=25 455 580 770]{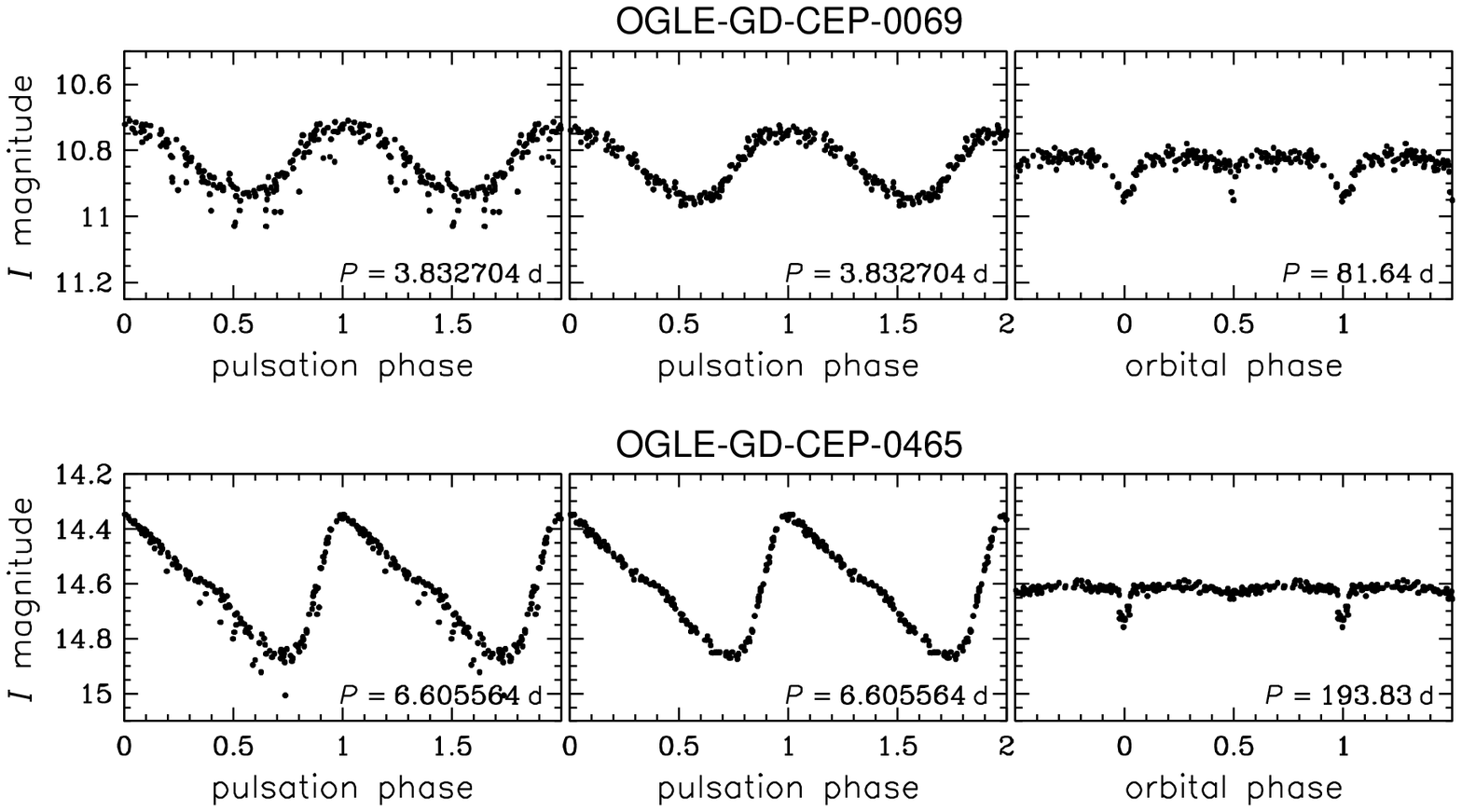}}
\FigCap{OGLE-GD-CEP-0069 {\it (top panel)} and OGLE-GD-CEP-0465 {\it
(bottom panel)} -- new candidates for eclipsing binary systems with a
classical Cepheid as one of the components. {\it Left panel} shows the
original data folded with the pulsation period. {\it Middle panel}
presents the pulsation light curve and the {\it right panel} --
eclipsing light curve.}
\end{figure}

\begin{figure}[htb]
\centerline{\includegraphics[width=11.7cm, bb=25 450 580 775]{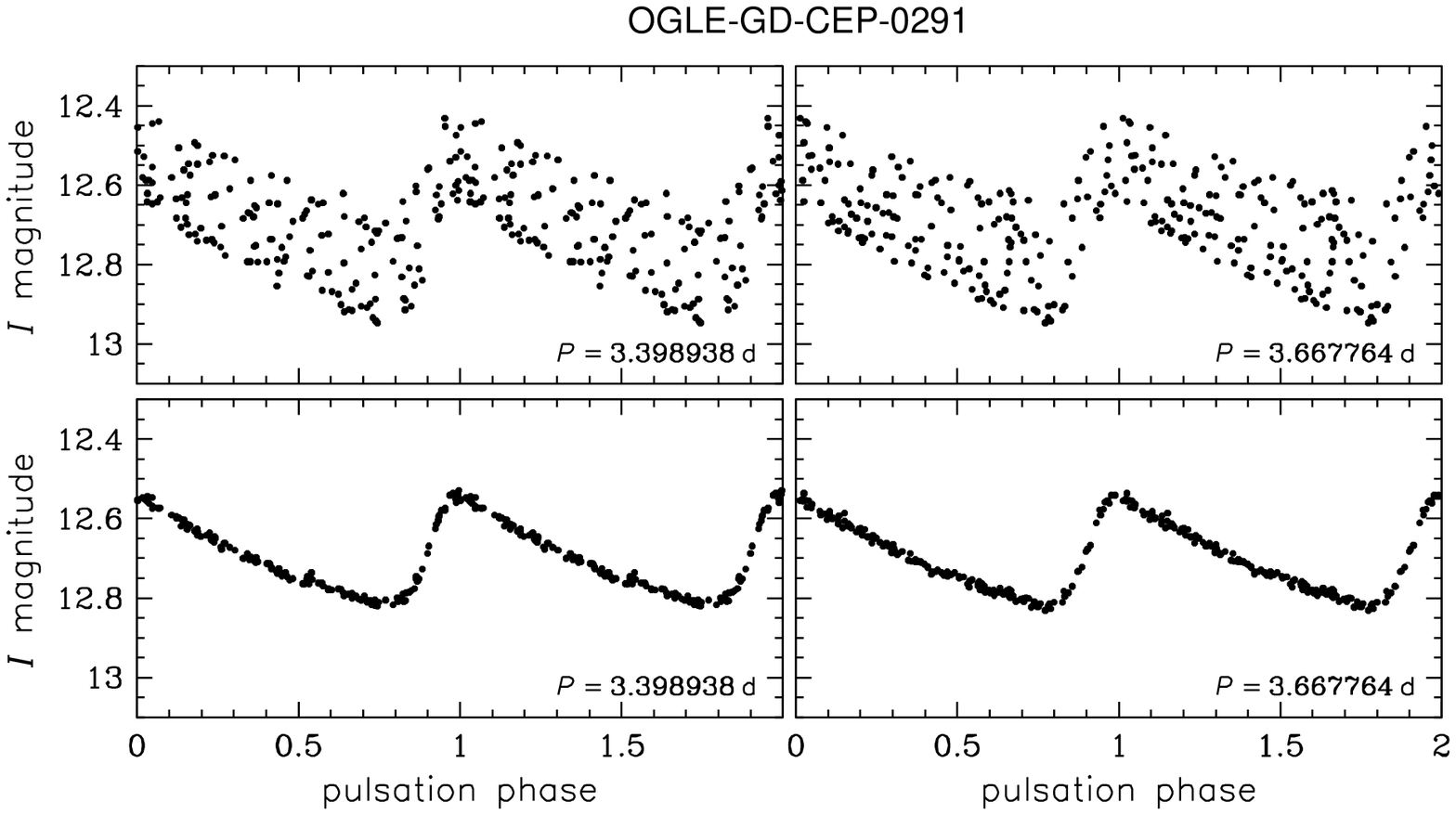}}
\FigCap{OGLE-GD-CEP-0291 -- candidate for a binary system containing two
classical Cepheids. {\it Top panel} shows the original data folded with
the pulsation period of each of the components. {\it Lower panel}
presents the pulsation light curve after subtracting other variability.}
\end{figure}
The third object -- OGLE-GD-CEP-0291 -- reveals light variations of a
Cepheid shape with two different periods: $P_1=3.398938\pm0.000018$~d and
$P_2=3.667764\pm0.000024$~d. Thus, it is very likely that this object is a
binary system containing two Cepheids pulsating in the fundamental
mode. Another possibility is a by-chance blend of two unrelated
Cepheids. However, this possibility seems to be unlikely -- also in this
case the stellar density of the field is not too high. Spectroscopic
follow-up will provide the status of this interesting object. Fig.~6 shows
the light curves of individual components of the system.

It is worth noticing that similar binary systems containing two Cepheids
were found in the LMC (Alcock \etal 1995, Udalski \etal 1999, Soszyñski
\etal 2008a). In one of these systems, OGLE-LMC-CEP-1718, Soszyñski \etal
(2008a) found eclipses and with follow-up spectroscopy from the Araucaria
project the system was solved by Gieren \etal (2014). Unfortunately, our
Galactic system does not reveal eclipses what indicates that the
inclination must be lower.

\Section{Discussion}
\vspace*{9pt}
OGLE Collection of Galactic Cepheids constitutes a new major OGLE dataset
of variable stars. After the OGLE Collection of Cepheids in the Magellanic
System (Soszyñski \etal 2015a, 2017a) this is the next important step in
studies of these very important objects of modern astrophysics.

The OGLE Collection more than doubles the number of known Galactic
classical Cepheids and considerably increases the number of other types of
these pulsators. The OGLE sample is very complete and not contaminated by
other variables. Additionally, precise and calibrated {\it VI} photometry
makes the OGLE dataset ideal for many scientific projects.

OGLE new Cepheids are located in a considerably larger volume of the Milky
Way, up to 20~kpc from the center, and cover practically the entire
disk. Based on this unique sample it has already been possible to prepare a
3-D picture of the Galaxy in young population as traced by Cepheids
(Skowron \etal 2018). This analysis showed the structure of the Galactic
disk like its warp, seen for the first time with individual stars having
accurate distance determination. One may also notice several clear
overdensities in the distribution of classical Cepheids in our Galaxy which
only roughly correspond to the present spiral arm pattern. However, these
overdensities can be modeled as a result of the past star formation
episodes in defined regions of the spiral arms and smeared up to now by the
Galactic and arm rotation (Skowron \etal 2018).

The OGLE Cepheids can also be excellent tracers of the Galactic
rotation. With the proper motions and radial velocities from the Gaia
mission (Gaia Collaboration \etal 2018) Mróz \etal (2018) have already presented the
rotation curve of the Milky Way reaching distances much farther from the
center than available for now. This study shows that the rotation curve is
basically flat up to the Galactic disk boundary. It can still be improved
when more radial velocities and proper motions of a significant part of the
OGLE Cepheids located far from the Galactic center become available.

OGLE Galactic Cepheids Collection can also be a new gold mine for studies
of properties of these pulsating stars. It provides new opportunities for
searching for additional low level pulsations. Many interesting additional
discoveries were done after releasing the previous parts of the OGLE
Cepheid Collection.

The OGLE Galactic Cepheid Collection released with this paper is an open
project. It will be updated in the future when additional fields monitored
during the GVS are ready for variable objects search and additional
samples of Galactic Cepheids are detected. Because classical Cepheids are
highly concentrated toward the Galactic plane and the most abundant regions
containing them are presented here, we do not expect a large increase of the
total number of new OGLE classical Cepheids in the subsequent updates of
our Collection. However, the number of other type Cepheids may
significantly increase, especially in the fields of the outer Galactic
bulge.

Additional update will occur when the search for faint Cepheids which did
not pass our preselection cut is finished. The ultimate goal of this OGLE
sub-project is to collect the vast majority of the Galactic Cepheids
observable from the Las Campanas Observatory.

\Acknow{We would like to thank M.~Kubiak, G.~Pietrzyñski,
£.~Wy\-rzy\-kow\-ski and M.~Pawlak for their contribution to the
collection of the OGLE photometric data presented in this paper. We
thank Z.~Ko³aczkowski and A.~Schwar\-zen\-berg-Czerny for providing
software used in this study.

The OGLE project has received funding from the Polish National Science
Centre grant MAESTRO 2014/14/A/ST9/00121 to AU. Support by the National
Science Centre, Poland, grant MAESTRO 2016/22/A/ST9/00009 to IS is also
acknowledged.

This research has made use of the International Variable Star Index (VSX)
database, operated at AAVSO, Cambridge, Massachusetts, USA.

This work has made use of data from the European Space Agency (ESA) mission
Gaia ({\it https://www.cosmos.esa.int/gaia}), processed by the Gaia Data
Processing and Analysis Consortium (DPAC, {\it
  https://www.cosmos.esa.int/web/gaia/dpac/ consortium}). Funding for the DPAC
has been provided by national institutions, in particular the institutions
participating in the Gaia Multilateral Agreement.  }

\end{document}